\documentclass[twocolumn,twocolappendix,tighten]{aastex631}
\usepackage{bm}%! To use \bm{} command (bold italic for the vector)
\usepackage{CJKutf8}%! To show the Japanese language.

\newcommand{\JLW}{J_{\rm LW}}
\newcommand{\Junit}{J_{21}}
\newcommand{\Jcrit}{J_{\rm crit}}
\newcommand{\fthree}{f_{\rm III}}
\newcommand{\K}{\rm K}
\newcommand{\tth}{t_{\rm th}}
\newcommand{\nth}{n_{\rm th}}

\newcommand{\msun}{M_\odot}
\newcommand{\msunyr}{M_\odot\,{\rm yr}^{-1}}
\newcommand{\nh}{n_{\rm H}}
\newcommand{\cc}{{\rm cm^{-3}}}

\graphicspath{{./}{src/}}

\begin{document}
\begin{CJK}{UTF8}{min}%! To show the Japanese language.

\shorttitle{Low-mass Pop III star by weak LW radiation}%[<=44 characters]
\title{Low-mass Pop III star formation due to the HD-cooling induced by weak Lyman-Werner radiation}

\correspondingauthor{Sho Nishijima}
\email{sho@nishijima.org}

\author[0009-0001-7178-4856]{Sho Nishijima}
\affiliation{Department of Astronomy, School of Science, University of Tokyo, Tokyo 113-0033, Japan}

\author[0000-0002-0786-7307]{Shingo Hirano}
\affiliation{Department of Astronomy, School of Science, University of Tokyo, Tokyo 113-0033, Japan}
\affiliation{Department of Applied Physics, Faculty of Engineering, Kanagawa University, Kanagawa 221-0802, Japan}

\author[0000-0001-8338-502X]{Hideyuki Umeda}
\affiliation{Department of Astronomy, School of Science, University of Tokyo, Tokyo 113-0033, Japan}

\begin{abstract}
Lyman-Werner (LW) radiation photodissociating molecular hydrogen (H$_2$) influences the thermal and dynamical evolution of the Population III (Pop III) star-forming gas cloud.
The effect of powerful LW radiation has been well investigated in the context of supermassive black hole formation in the early universe. 
However, the average intensity in the early universe is several orders of magnitude lower.
For a comprehensive study, we investigate the effects of LW radiation at $18$ different intensities, ranging from $\JLW/\Junit=0$ (no radiation) to $30$ (H-cooling cloud), on the primordial star-forming gas cloud obtained from a three-dimensional cosmological simulation.
The overall trend with increasing radiation intensity is a gradual increase in the gas cloud temperature, consistent with previous works.
Due to the HD-cooling, on the other hand, the dependence of gas cloud temperature on $\JLW$ deviates from the aforementioned increasing trend for a specific range of intensities ($\JLW/\Junit=0.025-0.09$).
In HD-cooling clouds, the temperature remained below $200$\,K during $10^5$\,yr after the first formation of the high-density region, maintaining a low accretion rate.
Finally, the HD-cooling clouds have only a low-mass dense core (above $10^8\,\cc$) with about $1-16\,\msun$, inside which a low-mass Pop III star with $\leq\!0.8\,\msun$ (so-called ``surviving star'') could form.
The upper limit of star formation efficiency $M_{\rm core}/M_{\rm vir, gas}$ significantly decreases from $10^{-3}$ to $10^{-5}$ as HD-cooling becomes effective.
This tendency indicates that, whereas the total gas mass in the host halo increases with the LW radiation intensity, the total Pop III stellar mass does not increase similarly.
\end{abstract}


%%%%%%%%%%%%%%%%%%%%%%%%%%%%%%%%%%%%%%%%%%%%%%%%%%
\section{Introduction} \label{sec:intro}
%%%%%%%%%%%%%%%%%%%%%%%%%%%%%%%%%%%%%%%%%%%%%%%%%%

The first stars, so-called Population III (Pop III) stars, formed from the primordial (metal-free) gas in the early Universe at $z = 30 - 20$ inside dark matter (DM) minihalo with $10^{5-6}\,\msun$ \citep{Tegmark1997}.
Previous theoretical and numerical works have uncovered the hierarchical formation process of Pop III stars, from the large-scale structure formation to the physical processes around the accreting protostar \citep[see][for a review]{KlessenGlover2023}.
The critical physical parameter of Pop III stars is the initial mass function (IMF) to evaluate their role in the formation and evolution of the first galaxies, which is one of the main observational targets of the James Webb Space Telescope (JWST).

Unlike nearby star-forming gas clouds where dust and metal are present, primordial star-forming gas clouds collapse by releasing the internal energy via molecular hydrogen (H$_2$) radiative cooling \citep[e.g.,][]{Abel2002, Bromm2002}.
The difference in the thermal evolution of gas clouds affects the stellar mass and the shape of IMF.
When the gas cloud becomes self-gravitationally unstable (Jeans unstable), the temperature of the primordial gas is around $200$\,K, whereas the temperature of the solar-metallicity gas drops to $10$\,K \citep[e.g.,][]{Omukai2005}.
Since the accretion rate, estimated by dividing the Jeans mass by the free-fall time, is proportional to $3/2$ power of the temperature ($\dot{M} \propto T^{3/2}$), the high temperature of the primordial gas is responsible for the increased mass accretion rate onto protostars and hence massive stellar mass ($\sim\!100\,\msun$) than the present-day case ($\sim\!1\,\msun$).

In the primordial gas, there are two other coolants besides H$_2$, and the thermal evolution and stellar mass will vary depending on which coolant is dominant.
One is the atomic hydrogen (H), which can act when H$_2$ formation is suppressed due to additional effects.
The H-cooling clouds can only be cooled to $8000$\,K and form more massive Pop III stars than the H$_2$-cooling clouds.
Such a situation is intensively investigated in scenarios that consider the formation of supermassive stars (intermediate-mass black holes) in the early universe as the origin of the high-$z$ quasars \citep[see][for reviews]{Woods2019, Inayoshi2020}.
Another is the hydrogen deuteride (HD), which becomes effective once the temperature of the primordial gas falls below $200$\,K, the condition under which HD formation becomes efficient.
The HD-cooling clouds can be cooled near the cosmic microwave background temperature floor $T_{\rm CMB} = 2.73(1+z)$\,K, thus forming low-mass Pop III stars \citep[e.g.,][]{Hosokawa2012b}.

For H-/HD-coolings to be effective, some external effects must be considered to change the gas cloud's physical properties.
One of the representative effects is H$_2$ photodissociation due to a radiation field incident from outside the cloud \citep{Omukai2001,OShea2008}.
If the external radiation intensity in Lyman-Werner (LW) bands is strong enough to photodissociate H$_2$ in the primordial gas cloud fully \citep[e.g., $\Jcrit = 100 - 1000$ in][]{Sugimura2014}, the gas cloud cools only with H-cooling, leading to the formation of supermassive stars.
Previous works investigated the supermassive star formation process to explore the formation scenario of seed black holes of high-$z$ quasars while considering such powerful radiation fields \citep[e.g.,][]{Dunn2018,ReganDownes2018b}.

The supermassive star formation under such critical radiation intensity is, on the whole, an extremely low-incidence event, comparable to the observed presence of high-$z$ quasars \citep[$\sim\!{\rm Gpc}^{-3}$,][]{Banados2016}.
The background radiation intensity during Pop III star formation is about a few orders of magnitude less than $\Jcrit$ \citep[e.g.,][]{Agarwal2012}.
In other words, most cases are associated with weak radiation intensities for the entire Pop III star formation event.
For a comprehensive understanding of the effect of external radiation on Pop III star formation and IMF, we have to consider the effect of such weak radiation intensities.
Recent studies have updated the Pop III star formation scenario inside the primordial cloud irradiated by external LW radiation, e.g., gas fragmentation and formation of less massive objects \citep{Regan2018,Suazo2019,Prole2023}, no correlation between the LW radiation intensity and host halo masses when considering the self-shielding effect \citep{Skinner2020}.

The next generation of telescopes is beginning to explore the early Universe, and a comprehensive understanding of the formation and evolutionary processes of first galaxies is becoming increasingly important.
We have performed Pop III star formation simulations in all possible ranges of radiation intensity, from no radiation field to $\Jcrit$, with $18$ different radiation intensities.
We skip the calculation of regions with a density of $\nh = 10^8\,\cc$ and above, and thus successfully calculate the accretion process from the halo to the high-density regions for 100,000 years after the high-density region formation.
We investigate how the various thermal evolution of gas clouds, which appear according to the external radiation intensity used as a parameter, affects the accretion growth of dense cores that host Pop III stars inside.

This paper is organized as follows.
Section~\ref{sec:method} summarizes the numerical methodology of the cosmological simulations.
Section~\ref{sec:results} shows the simulation results under different LW radiation intensities.
Section~\ref{sec:dis} discusses why HD-cooling is effective in gas clouds exposed to LW radiation and determines the Pop III stellar mass and star formation efficiency.
Section~\ref{sec:sum} finally summarizes the main conclusions of this study.

%%%%%%%%%%%%%%%%%%%%%%%%%%%%%%%%%%%%%%%%%%%%%%%%%%%%%%%%%%%%
\begin{deluxetable*}{rccccrccrcc}[t]
\tablecaption{Model Parameters and Simulation Results}
\tablecolumns{11}
\tablenum{1}
\tablewidth{0pt}
\tablehead{
  \colhead{1} &
  \colhead{2} &
  \colhead{3} &
  \colhead{4} &
  \colhead{5} &
  \colhead{6} &
  \colhead{7} &
  \colhead{8} &
  \colhead{9} &
  \colhead{10} &
  \colhead{11} \\
  \colhead{$\JLW$} &
  \colhead{$z$} &
  \colhead{$R_{\rm vir}$} &
  \colhead{$M_{\rm vir}$} &
  \colhead{$M_{\rm vir, gas}$} &
  \colhead{$t_{\rm col} / t_{\rm ff}$} &
  \colhead{HD} &
  \colhead{HD} &
  \colhead{$M_{\rm core}$} &
  \colhead{$\fthree$} &
  \colhead{$\dot{M}$} \\
  \colhead{($\Junit$)} &
  \colhead{} &
  \colhead{(pc)} &
  \colhead{$(\msun)$} &
  \colhead{$(\msun)$} &
  \colhead{} &
  \colhead{$\tth=0$\,yr} &
  \colhead{$\tth=10^5$\,yr} &
  \colhead{$(\msun)$} &
  \colhead{}&
  \colhead{$(\msunyr)$}
}
\startdata
 0.000 & $24.88$ & $125.9$ & $8.74 \times 10 ^ {5}$ & $1.29 \times 10^{5}$ &  5.63 & Y & Y &     $3.86$         & $2.99 \times 10^{-5}$ & $9.54 \times 10^{-4}$ \\
\hline
 0.003 & $23.65$ & $100.0$ & $7.34 \times 10 ^ {5}$ & $1.05 \times 10^{5}$ &  4.48 & N & N &   $381\hspace{1.2em}$ & $1.31 \times 10^{-2}$ & $5.35 \times 10^{-2}$ \\
 0.010 & $23.04$ & $141.3$ & $1.14 \times 10 ^ {6}$ & $1.74 \times 10^{5}$ &  4.01 & N & N &   $819\hspace{1.2em}$ & $4.72 \times 10^{-3}$ & $9.88 \times 10^{-2}$ \\
 0.015 & $22.94$ & $158.5$ & $1.42 \times 10 ^ {6}$ & $2.19 \times 10^{5}$ &  2.15 & N & N &   $526\hspace{1.2em}$ & $2.41 \times 10^{-3}$ & $1.50 \times 10^{-2}$ \\
 0.020 & $22.54$ & $158.5$ & $1.54 \times 10 ^ {6}$ & $2.38 \times 10^{5}$ &  3.26 & Y & N &   $263\hspace{1.2em}$ & $1.11 \times 10^{-3}$ & $5.00 \times 10^{-3}$ \\
\hline
 0.025 & $21.96$ & $177.8$ & $1.96 \times 10 ^ {6}$ & $2.97 \times 10^{5}$ &  5.40 & Y & Y &     $6.07$        & $2.05 \times 10^{-5}$ & $4.34 \times 10^{-4}$ \\
 0.030 & $20.44$ & $177.8$ & $1.86 \times 10 ^ {6}$ & $2.80 \times 10^{5}$ & 10.92 & Y & Y &     $1.15$         & $4.09 \times 10^{-6}$ & $2.99 \times 10^{-5}$ \\
 0.040 & $20.67$ & $223.9$ & $2.94 \times 10 ^ {6}$ & $4.51 \times 10^{5}$ &  6.65 & Y & Y &     $4.47$         & $9.90 \times 10^{-6}$ & $4.34 \times 10^{-4}$ \\
 0.050 & $20.44$ & $223.9$ & $3.07 \times 10 ^ {6}$ & $4.71 \times 10^{5}$ &  5.50 & Y & Y &    $15.92$ & $3.38 \times 10^{-5}$ & $1.03 \times 10^{-3}$ \\
 0.065 & $19.91$ & $251.2$ & $3.71 \times 10 ^ {6}$ & $5.72 \times 10^{5}$ &  3.76 & Y & Y &     $3.10$         & $5.42 \times 10^{-6}$ & $2.42 \times 10^{-3}$ \\
 0.080 & $19.20$ & $281.8$ & $4.80 \times 10 ^ {6}$ & $7.39 \times 10^{5}$ &  4.02 & Y & Y &     $3.16$         & $4.28 \times 10^{-6}$ & $4.48 \times 10^{-4}$ \\
 0.090 & $17.67$ & $354.8$ & $8.23 \times 10 ^ {6}$ & $1.27 \times 10^{6}$ &  9.35 & Y & Y &     $2.23$         & $1.76 \times 10^{-6}$ & $1.66 \times 10^{-4}$ \\
\hline
 0.100 & $15.97$ & $446.7$ & $1.09 \times 10 ^ {7}$ & $1.68 \times 10^{6}$ &  5.16 & Y & N &   $435\hspace{1.2em}$ & $2.60 \times 10^{-4}$ & $4.66 \times 10^{-3}$ \\
 0.300 & $14.87$ & $562.3$ & $1.83 \times 10 ^ {7}$ & $2.82 \times 10^{6}$ &  3.22 & Y & N &   $737\hspace{1.2em}$ & $2.62 \times 10^{-4}$ & $2.79 \times 10^{-2}$ \\
 1.000 & $14.13$ & $631.0$ & $2.20 \times 10 ^ {7}$ & $3.41 \times 10^{6}$ &  3.24 & N & N &   $715\hspace{1.2em}$ & $2.10 \times 10^{-4}$ & $1.64 \times 10^{-2}$ \\
 3.000 & $13.57$ & $631.0$ & $2.44 \times 10 ^ {7}$ & $3.78 \times 10^{6}$ &  3.02 & Y & N &   $349\hspace{1.2em}$ & $9.24 \times 10^{-5}$ & $1.20 \times 10^{-2}$ \\
10.000 & $13.35$ & $707.9$ & $2.66 \times 10 ^ {7}$ & $4.13 \times 10^{6}$ &  2.31 & Y & N &   $611\hspace{1.2em}$ & $1.48 \times 10^{-4}$ & $1.82 \times 10^{-2}$ \\
\hline
30.000 & $13.08$ & $707.9$ & $2.76 \times 10 ^ {7}$ & $4.27 \times 10^{6}$ &  3.82 & N & N & $7.09\times10^4$ & $1.66 \times 10^{-2}$ & $7.20 \times 10^{+0}$ \\
\enddata
%\tablenotetext{a}{At exposure start.}
\tablecomments{
Column (1): radiation intensity at the Lyman-Werner bands in units of $\Junit = 10^{-21}$\,erg\,s$^{-1}$\,cm$^{-2}$\,Hz$^{-1}$\,sr$^{-1}$.
Column (2): redshift when the maximum gas number density firstly reaches $10^8\,\cc$ ($\tth = 0$\,yr).
Columns (3-5): radius, total mass, and gas mass at the virial scale where the total matter density exceeds $200$ times the cosmic average.
Column (6): ratio of the collapse and free-fall timescales averaged at $\nh = 10^{3}-10^{5}\,\cc$.
Columns (7) and (8): whether HD-cooling is enabled in the gas cloud (whether the abundance ratio of H$_2$ and HD molecules overcomes the critical value $f_{\rm HD}/f_{\rm H_2} \geq 10^{-3}$) at $\tth = 0$\,yr and $10^5$\,yr.
Column (9): core mass where $\nh \geq 10^8\,\cc$.
Column (10): star formation efficiency, $\fthree = M_{\rm core}/M_{\rm vir, gas}$.
Column (11): mass accretion rate at the Jeans radius when $\tth = 0$\,yr.
We end simulations at $\tth = 10^5$\,yr, except for $\JLW/\Junit = 30$ model, for which we stopped the calculation at $\tth = 5\times10^4$\,yr.
The horizontal lines distinguish the five ways gas clouds are classified according to their thermal evolution (R1-R5; see Section~\ref{sec:results:cloud}).
}
\label{tab:t1}
\end{deluxetable*}
%%%%%%%%%%%%%%%%%%%%%%%%%%%%%%%%%%%%%%%%%%%%%%%%%%%%%%%%%%%%

%%%%%%%%%%%%%%%%%%%%%%%%%%%%%%%%%%%%%%%%%%%%%%%%%%%%%%%%%%%%
\begin{figure*}
\begin{center}
\includegraphics[width=1.0\linewidth]{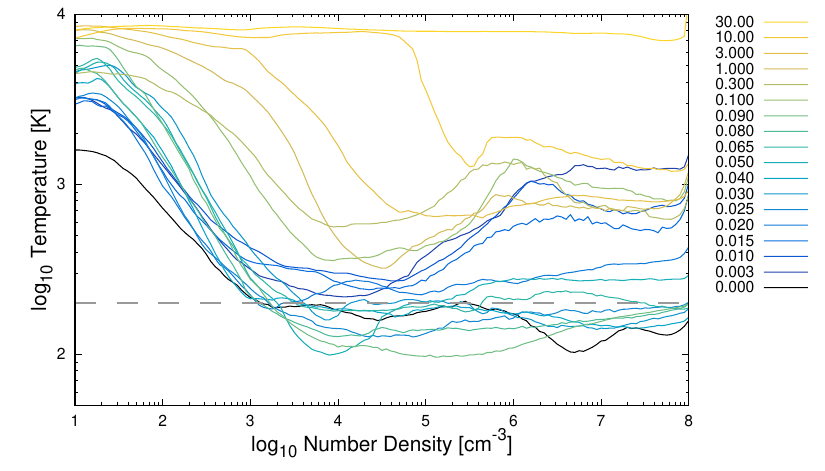}
\end{center}
\caption{
Thermal distributions of primordial star-forming gas clouds at $\tth = 10^5$\,yr as a function of the gas density.
Each color represents the different intensities of LW radiation ($\JLW/\Junit$).
The gray dashed line represents $200$\,K.
}
\label{fig:f1}
\end{figure*}
%%%%%%%%%%%%%%%%%%%%%%%%%%%%%%%%%%%%%%%%%%%%%%%%%%%%%%%%%%%%

%%%%%%%%%%%%%%%%%%%%%%%%%%%%%%%%%%%%%%%%%%%%%%%%%%
\section{Methods} \label{sec:method}
%%%%%%%%%%%%%%%%%%%%%%%%%%%%%%%%%%%%%%%%%%%%%%%%%%

We perform three-dimensional cosmological $N$-body / hydrodynamical simulations of the Pop III star formation under different external Lyman-Werner (LW) background radiation intensities.
We use a hierarchical zoom-in technique to achieve a sufficiently high spatial resolution to follow the formation of the star-forming gas cloud within the dark matter (mini)halo.
We adopt a stiff equation of state (EOS) technique to follow the long-term dynamical and thermal evolution of the star-forming gas cloud.

We generate a cosmological initial condition using the publicly available code {\sc music} \citep{Hahn2011}.
The parent cosmological simulation has a volume of $1\,h^{-1}$\,comoving\,Mpc (cMpc) on a side.
We insert a series of nested refinement regions, reaching the maximum refined region inside a volume of $0.15\,h^{-1}$\,cMpc on a side.
The particle masses of dark matter and gas components in the maximally refined region are $m_{\rm DM} = 12.4\,\msun$ and $m_{\rm gas} = 2.3\,\msun$, respectively.
We adopt the standard $\Lambda$-Cold Dark Matter ($\Lambda$-CDM) cosmology with cosmological parameters: matter density $\Omega_{\rm m} = 0.3086$, baryon density $\Omega_{\rm b} = 0.04825$, dark energy density $\Omega_{\Lambda} = 0.6914$ in units of the critical density, Hubble constant of $h = 0.6777$, normalization of the density fluctuation amplitude $\sigma_8 = 0.8288$, and primordial index $n_{\rm s} = 0.961$ \citep{PLANCK13XVI}.
The initial ionization fraction is $x_{\rm e} = 2.737 \times 10^{-4}$ \citep{seager99,seager00,wong08}.

The cosmological simulations are performed by using the parallel $N$-body / Smoothed Particle Hydrodynamics (SPH) code {\sc gadget-2} \citep{Springel2005}, suitably modified for the primordial star formation case \citep{Hirano2018} with solving chemical rate equations for 14 primordial species \citep[e$^{-}$, H, H$^{+}$, H$^{-}$, He, He$^{+}$, He$^{++}$, H$_2$, H$^{+}_2$, D, D$^{+}$, HD, HD$^{+}$, HD$^{-}$;][]{Yoshida2006,Yoshida2007}.
We employ a hierarchical refinement technique to follow the gas cloud collapse with the refinement criterion that the local Jeans length is always resolved by $15$ times the local smoothing length by progressively increasing the spatial resolution using the particle-splitting technique \citep{Kitsionas2002}.
The minimum mass of baryon particles becomes $m_{\rm gas,min} = 8.05 \times 10^{-3}\,\msun$.

This study considers the effect of external radiation in the LW bands, which photodissociates the important coolants, H$_2$ and HD molecules.
We assume a blackbody spectrum of Pop II stars with the effective temperature of $10^4$\,K as the light source.
The collapsing gas cloud increases its central density and finally becomes optically thick.
After that, the opaque shell surrounds the further collapsing region, and the abundance of both coolants can recover because the external radiation is consumed by photodissociating the surrounding gas (self-shielding mechanism).
We adopt the self-shielding functions \citep{Wolcott-GreenHaiman2011, Wolcott-Green2011} for H$_2$ and HD molecules.
For each SPH particle, we calculate the column densities along six directions ($\pm X$, $\pm Y$, $\pm Z$) to account for the directional dependence of the self-shielding effect \citep[six-ray approximation;][]{Yoshida2008}.

We follow cosmological structure formation from redshift $z = 99$ to $39$.
Then, we restart cosmological simulations by adding uniform LW radiations from $z = 39$.\footnote{The switching redshift ($z = 39$) is too large for the background radiation field from the Pop II star to be present. This study investigates the effect of subtle differences in LW radiation intensity on the thermal evolution of primordial gas clouds. A more realistic investigation that considers the distribution of the background radiation field during the formation of gas clouds is left to future work.}
The model parameter is the LW radiation intensity $\JLW$ that is the intensity at the LW bands normalized in units of $\Junit = 10^{-21}$\,erg\,s$^{-1}$\,cm$^{-2}$\,Hz$^{-1}$\,sr$^{-1}$.
We adopt $18$ models with $\JLW/\Junit=0-30$ (Table~\ref{tab:t1}).
The model with the maximum intensity $\JLW/\Junit = 30$ corresponds to the atomic-cooling halo (ACH) case.
We specifically examine the parameter dependence in more detail by taking nine parameters between $\JLW/\Junit = 0.02 - 0.09$.
We calculate the gravitational collapse of the star-forming gas cloud until the gas number density first reaches $\nh = \rho / m_{\rm H} = 10^8\,\cc$.

Finally, we study the long-term evolution of the star-forming gas cloud.
To accelerate the evolution, we adopt a stiff EOS technique with a threshold density $\nth = 10^8\,\cc$ above which the gravitational collapse is artificially prohibited \citep{Hirano2017b}.
This study assumes the dense region with $\nh \geq \nth$ (``core'') as the host site where the Pop III stars form in the interior.
We analyze the formation and evolution of cores (mass and number) in simulations with different $\JLW/\Junit$.
We continue hydrodynamical simulations for $10^5$\,yr \citep[typical timescale of the Pop III star accretion phase, e.g., Figure~1 in][]{Hirano2017b} after the gas density first reached $\nth = 10^8\,\cc$.
For the case with $\JLW/\Junit = 30$, we stop the simulation at $5 \times 10^4$\,yr because of the enormous computational cost and because it is not the intermediate radiation intensity, which is the main objective of this study.

%%%%%%%%%%%%%%%%%%%%%%%%%%%%%%%%%%%%%%%%%%%%%%%%%%
\section{Results} \label{sec:results}
%%%%%%%%%%%%%%%%%%%%%%%%%%%%%%%%%%%%%%%%%%%%%%%%%%

Figure~\ref{fig:f1} overviews the thermal evolution of primordial star-forming gas clouds irradiated by $18$ different LW radiation intensities.
The gas temperature increases with increasing radiation intensities $\JLW/\Junit$ due to the increase of H$_2$ photodissociation rate.
In particular, in the case of the highest intensity $\JLW/\Junit = 30$, the self-shielding effect is not effective enough to maintain high H$_2$ photodissociation rates, resulting in an isothermally contracting H-cooling gas cloud at $8000$\,K.
With the intermediate intensities $\JLW/\Junit = 0.025 - 0.09$, on the other hand, HD-cooling works, and the gas temperature in the high-density region remained below $200$\,K until the end of simulations.
In HD-cooling clouds, the mass of the high-density region decreases by two orders of magnitude compared to the others (Table~\ref{tab:t1}), which could impact the Pop III stellar mass born in their interiors.
The following parts explain how temperature evolution varies qualitatively with radiation intensity.

%%%%%%%%%%%%%%%%%%%%%%%%%%%%%%%%%%%%%%%%%%%%%%%%%%%%%%%%%%%%
\begin{figure}
\begin{center}
\includegraphics[width=\linewidth]{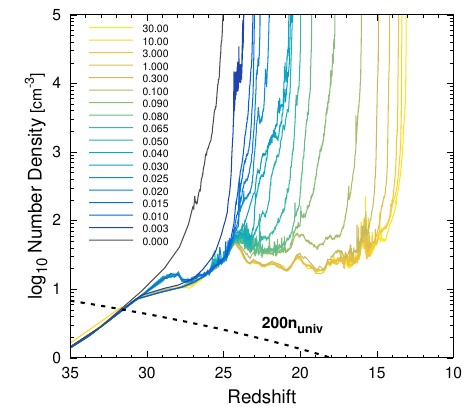}
\end{center}
\caption{
Time evolution of the maximum density of the collapsing gas cloud as a function of redshift.
The dashed line indicates $200$ times the cosmic average density, $200\,n_{\rm H,univ}$, which defines the virial scale.
}
\label{fig:f2}
\end{figure}
%%%%%%%%%%%%%%%%%%%%%%%%%%%%%%%%%%%%%%%%%%%%%%%%%%%%%%%%%%%%

%%%%%%%%%%%%%%%%%%%%%%%%%%%%%%%%%%%%%%%%%%%%%%%%%%%%%%%%%%%%
\begin{figure*}
\begin{center}
\includegraphics[width=1.0\linewidth]{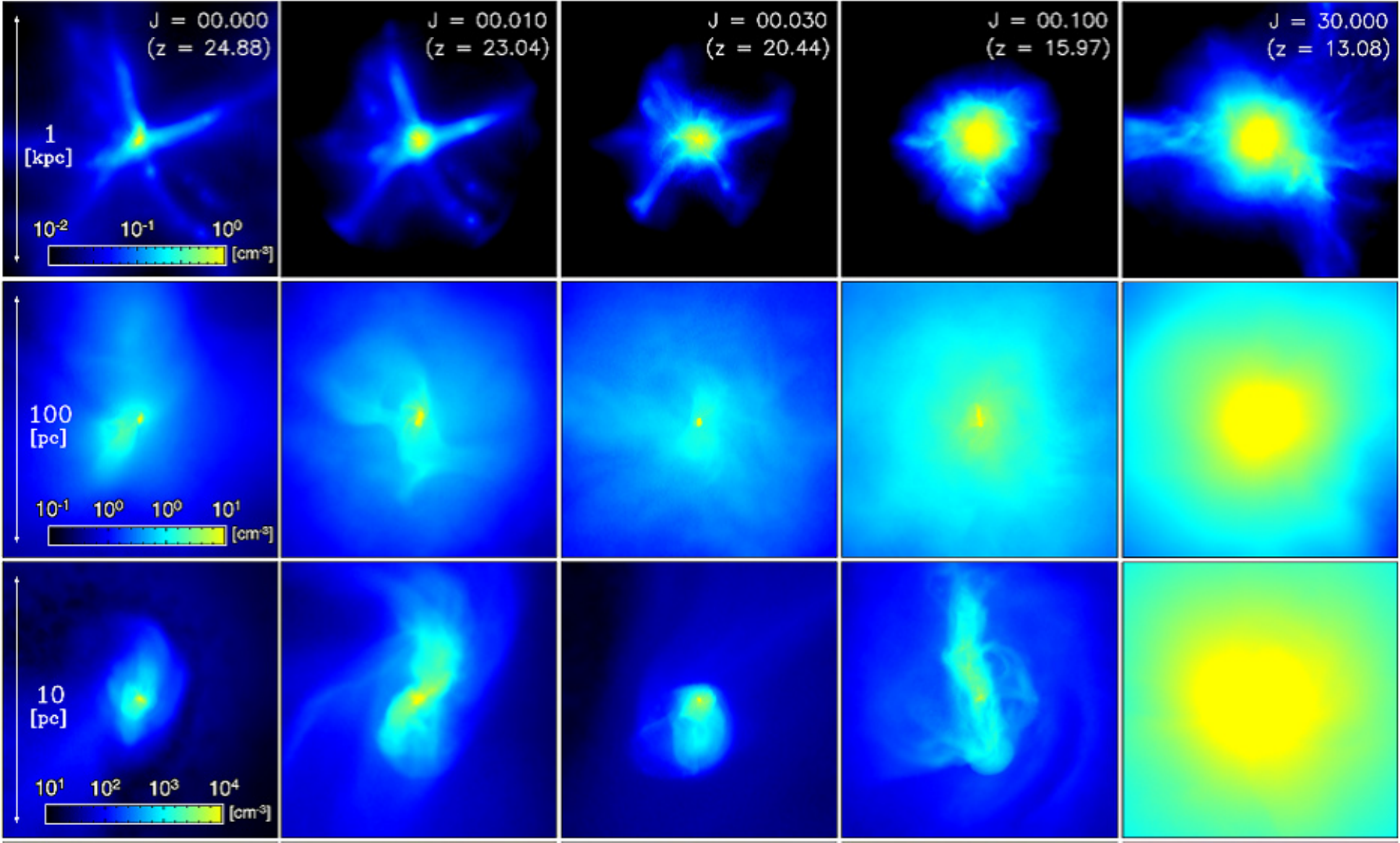}
\end{center}
\caption{
Gas density distribution around the collapsing center of the primordial gas cloud for the cases with $\JLW/\Junit = 0$, $0.01$, $0.03$, $0.10$, and $30$ when $\tth = 0$\,yr.
The box sizes are $1$\,kpc (top panels), $100$\,pc (middle), $10$\,pc (bottom) on a side, respectively.
}
\label{fig:f3}
\end{figure*}
%%%%%%%%%%%%%%%%%%%%%%%%%%%%%%%%%%%%%%%%%%%%%%%%%%%%%%%%%%%%

%%%%%%%%%%%%%%%%%%%%%%%%%%%%%%%%%%%%%%%%%%%%%%%%%%%%%%%%%%%%
\subsection{Delay of the halo formation} \label{sec:results:halo}
%%%%%%%%%%%%%%%%%%%%%%%%%%%%%%%%%%%%%%%%%%%%%%%%%%%%%%%%%%%%

First, we discuss the effects of different LW radiation intensities on the host halo properties.
Columns~2-5 in Table~\ref{tab:t1} summarize the physical properties at the host halo scale.

As $\JLW$ increases from $\JLW/\Junit = 0$ to $30$, the formation epoch delays from $z = 24.88$ to $13.08$ and host halo mass increases from $M_{\rm vir} = 8.74\times10^5\,\msun$ to $2.76\times 10^7\,\msun$.
Figure~\ref{fig:f2} shows the time evolution of the maximum density of the collapsing gas inside the host halo.
In models with $\JLW/\Junit > 0$, the gas contraction stalls around $\nh \sim 10\,\cc$ at which the gas temperature changes from an increase due to adiabatic compression to a decrease due to radiative cooling (Figure~\ref{fig:f1}).
Because the H$_2$ photodissociation rate increases proportionately to the LW radiation intensity, a larger halo mass is necessary to achieve the H$_2$ formation rate above the H$_2$ photodissociation rate.

The LW radiation intensities, which determine the thermal evolution of the collapsing clouds (Figure~\ref{fig:f1}), also affect the structure of the collapsing gas clouds inside host halos (Figure~\ref{fig:f3}).
The radius of halo increases from $R_{\rm vir} = 126$\,pc to $708$\,pc with the LW radiation intensity (Table~\ref{tab:t1}).
In $1$\,kpc square region (top panels of Figure~\ref{fig:f3}), all halos show a spherical density distribution, which is the gas cloud's structure during the increasing temperature phase due to adiabatic compression.
The $100$\,pc square region (middle) shows the gas density distribution inside the halo.
The three left panels show a structure that deviates from spherical symmetry due to the rapid temperature drop caused by H$_2$-cooling.
On the other hand, the two right panels show a broadened spherically symmetric structure due to the strong LW radiation inhibiting H$_2$ formation.
At the $10$\,pc (bottom) scale, $\JLW/\Junit = 0.1$ model shows a large filament, where the gas temperature decreases at $100\,\cc$ (Figure~\ref{fig:f1}), and the structure of collapsing gas changes.
On the other hand, the cloud of $\JLW/\Junit = 30$ model isothermally collapses due to H-cooling ($\sim\!8000$\,K) and thus retains a spherically symmetric structure up to this scale.
Although spherically symmetric at the DM halo scale, the structure of the dense gas cloud varies from spherically symmetric to filamentary due to different temperature evolution in its interior.

%%%%%%%%%%%%%%%%%%%%%%%%%%%%%%%%%%%%%%%%%%%%%%%%%%%%%%%%%%%%
\begin{figure*}
\begin{center}
\includegraphics[width=1.0\linewidth]{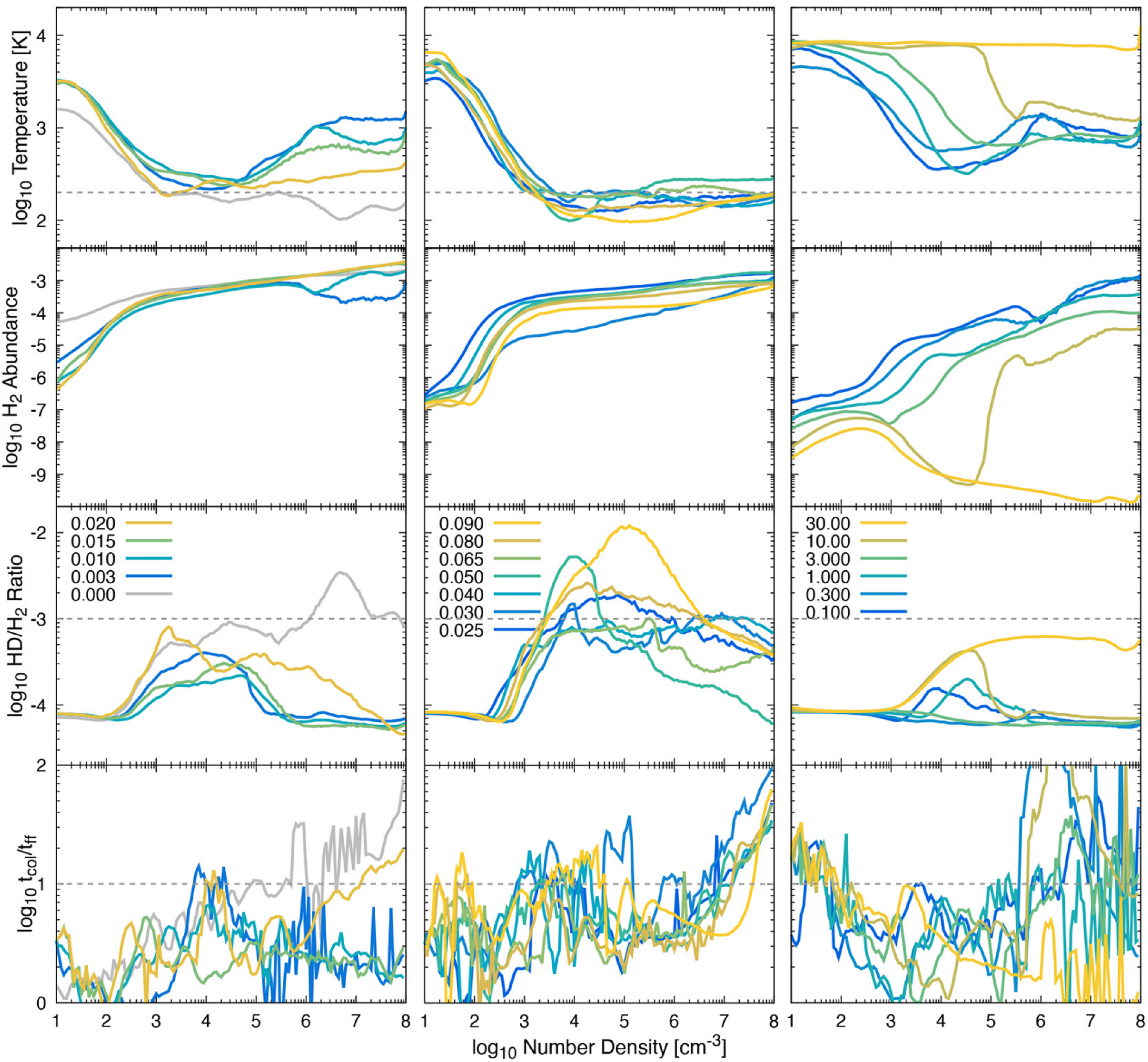}
\end{center}
\caption{
Thermal properties of the gas cloud as a function of gas number density: temperature, H$_2$ abundance, the ratio of H$_2$ and HD abundances, and the ratio of the collapse and free-fall timescales ${t_{\rm col}/t_{\rm ff}}$ (from top to bottom panels).
The top three rows are calculated from data at $\tth = 10^5$\,yr, and the bottom row is calculated using the maximum density evolution during the cloud collapse until $\tth = 0$\,yr.
Line colors represent the model parameter, radiation intensities $\JLW/\Junit$.
We classify the models into three groups, using the intensity range with which the HD-cooling becomes effective as the boundary: (left) $\JLW/\Junit = 0-0.02$, (center) $\JLW/\Junit = 0.025-0.09$, and (right) $\JLW/\Junit = 0.1-30$.  
On the left panels, HD-cooling is enabled exceptionally on models without LW radiation ($\JLW/\Junit = 0$).
}
\label{fig:f4}
\end{figure*}
%%%%%%%%%%%%%%%%%%%%%%%%%%%%%%%%%%%%%%%%%%%%%%%%%%%%%%%%%%%%

%%%%%%%%%%%%%%%%%%%%%%%%%%%%%%%%%%%%%%%%%%%%%%%%%%%%%%%%%%%%
\begin{figure*}
\begin{center}
\includegraphics[width=1.0\linewidth]{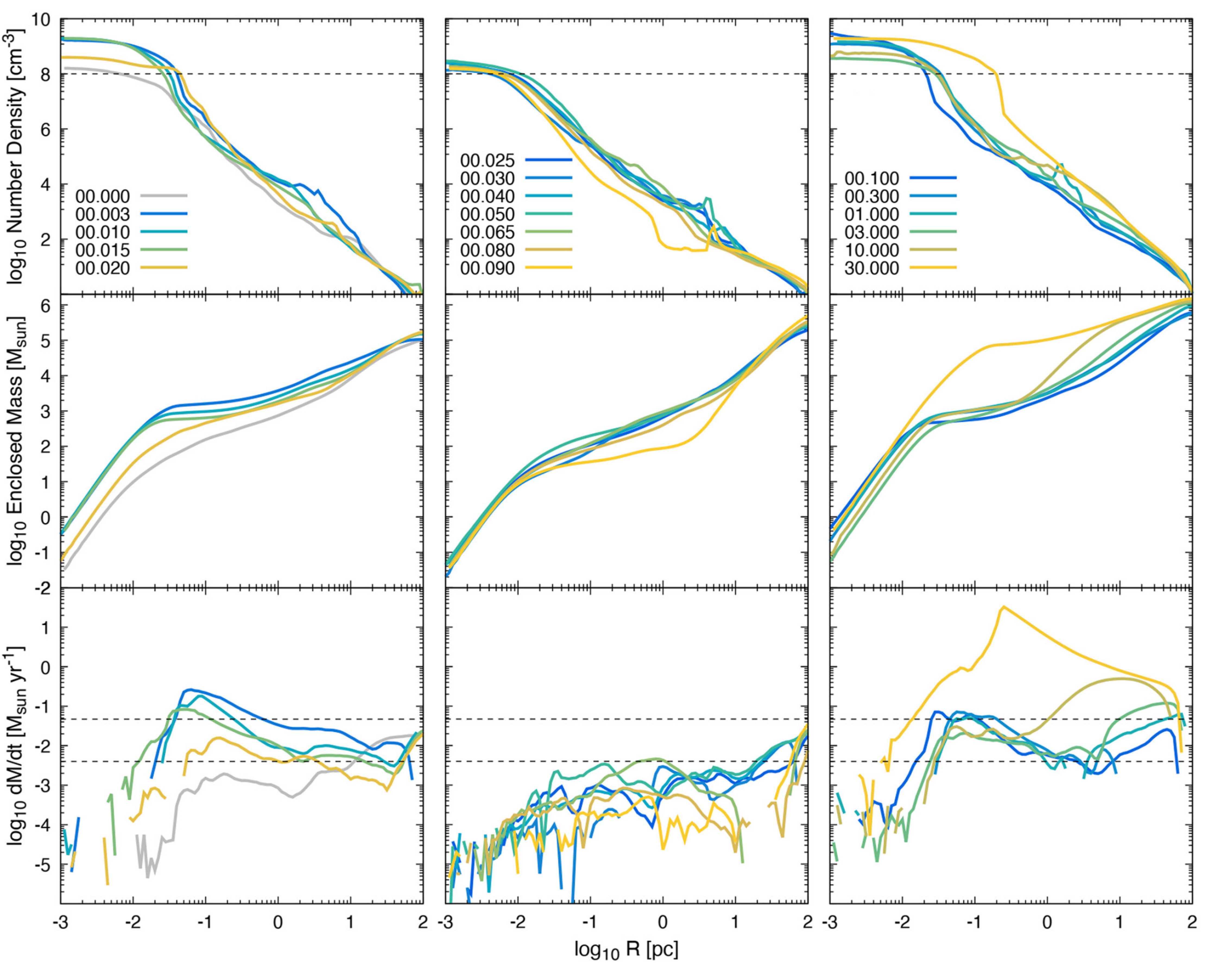}
\end{center}
\caption{
Radial profiles of (panel a) gas number density, (b) enclosed gas mass, and (c) gas accretion rate at $\tth = 10^5$\,yr.
We classify models into three groups based on parameter $\JLW/\Junit$ as same as Figure~\ref{fig:f4}.
The colors represent the model parameter $\JLW/\Junit = 0 - 30$.
The dashed lines in the top panels show $\nth = 10^{8}\,\cc$, which is the maximum resolution density of our simulation.
The dashed lines in the bottom panels show two critical values of the mass accretion rates, $4\times10^{-3}$ and $4.7\times10^{-2}\,\msunyr$, for the protostellar radiative feedback. 
}
\label{fig:f5}
\end{figure*}
%%%%%%%%%%%%%%%%%%%%%%%%%%%%%%%%%%%%%%%%%%%%%%%%%%%%%%%%%%%%

%%%%%%%%%%%%%%%%%%%%%%%%%%%%%%%%%%%%%%%%%%%%%%%%%%%%%%%%%%%%
\subsection{Various thermal evolution of gas cloud} \label{sec:results:cloud}
%%%%%%%%%%%%%%%%%%%%%%%%%%%%%%%%%%%%%%%%%%%%%%%%%%%%%%%%%%%%

The thermal evolution of gas clouds does not vary monotonically with increasing LW radiation intensity (Figure~\ref{fig:f1}).
We categorize 18 models into three types depending on the dominant chemical species contributing to the radiative cooling of the gas cloud: (a) HD-, (b) H$_2$-, and (c) H-cooling clouds.
HD-cooling cloud is cooled by deuterium hydrogen (HD), provided that the ratio of HD to H$_2$ is above the critical value of $f_{\rm HD}/f_{\rm H_2} = 10^{-3}$ \citep{Ripamonti2007}.
H-cooling cloud is cooled by atomic hydrogen (H), in which the H$_2$ photodissociation completely inhibits H$_2$ formation due to the intense LW radiation.
H$_2$-cooling cloud is the remaining case, where HD-cooling does not work and the internal energy of the gas cloud is lost mainly through H$_2$-cooling.

Following the above definition, we have classified $18$ models in five ranges with $\JLW/\Junit$ as a parameter:
\begin{itemize}
\item[(R1)] HD-cooling clouds for $\JLW/\Junit = 0$,
\item[(R2)] H$_2$ for $\JLW/\Junit = 0.003 - 0.02$,
\item[(R3)] HD for $\JLW/\Junit = 0.025 - 0.09$,
\item[(R4)] H$_2$ for $\JLW/\Junit = 0.1 - 10$, and
\item[(R5)] H for $\JLW/\Junit = 30$.
\end{itemize}
Note that these classifications are based on results at the end of the simulations ($\tth = 10^5$\,yr).

To identify differences in the thermal evolution of the gas clouds with the main coolant, Figure~\ref{fig:f4} divides $18$ models into three portions according to the range of $\JLW/\Junit$.
The left panels of Figure~\ref{fig:f4} show R1 and R2 models.
For the gas cloud without external radiation (R1; $\JLW/\Junit = 0$), the gas temperature became lower than $\sim\!200$\,K, where HD-cooling could be efficient.
The abundance ratio certainly exceeds the critical value for HD-cooling when the gas number density is above $10^6\,\cc$.
We classify the R1 model as HD-cooling cloud.

On the other hand, R2 models ($\JLW/\Junit = 0.003 - 0.02$) show relatively high-temperature evolution.
H$_2$ photodissociation is effective in the low-density region ($\nh < 10^{3}\,\cc$), reducing H$_2$ abundance, but as density increases, H$_2$ photodissociation becomes ineffective due to the self-shielding effect.
The thermal evolution of the gas cloud is driven by H$_2$-cooling, so the temperature of the gas cloud does not fall below $200$\,K.
We classify R2 models as H$_2$-cooling clouds.

The central panels of Figure~\ref{fig:f4} show the models with intermediate (``weak'') LW radiation (R3; $\JLW/\Junit = 0.025 - 0.090$).
The abundance ratio is higher than models in other panels and above the critical value required for HD-cooling.
As a result, the gas temperature remains below $200$\,K until the end of simulations.
We classify R3 models as HD-cooling clouds.
We will discuss why HD-cooling is enabled for R3 models in Section~\ref{sec:dis:origin}.

The right panels of Figure~\ref{fig:f4} show the remaining models irradiated by higher intensities.
In these cases, the high photodissociation rate counteracts the self-shielding effect, so the gas clouds have a low H$_2$ abundance up to a high density.
R4 models ($\JLW/\Junit = 0.1 - 10$) show the thermal evolution on the high-temperature side.
H$_2$ formation does not proceed until the density increase causes the star formation rate to exceed the photodissociation rate, and the gas cloud experiences a significant temperature drop.
This temperature gap produces the filamentary structure shown in Figure~\ref{fig:f3} \citep[see also Section~3 in][]{Hirano2023}.
Eventually, H$_2$-cooling becomes effective in these gas clouds.
We classify R4 models as HD-cooling clouds.

R5 model ($\JLW/\Junit = 30$) shows the isothermal collapse at $8000$\,K.
H$_2$ photodissociation rate is sufficiently high to suppress H$_2$ formation.
Only H-cooling cools the gas cloud and drives the cloud contraction while maintaining a high temperature \citep[zone of no return;][]{Inayoshi2012}.
H$_2$ formation can not proceed until the end of the simulation.
We classify the R5 model as the H-cooling cloud.

%%%%%%%%%%%%%%%%%%%%%%%%%%%%%%%%%%%%%%%%%%%%%%%%%%%%%%%%%%%%
\begin{figure}
\begin{center}
\includegraphics[width=1.0\linewidth]{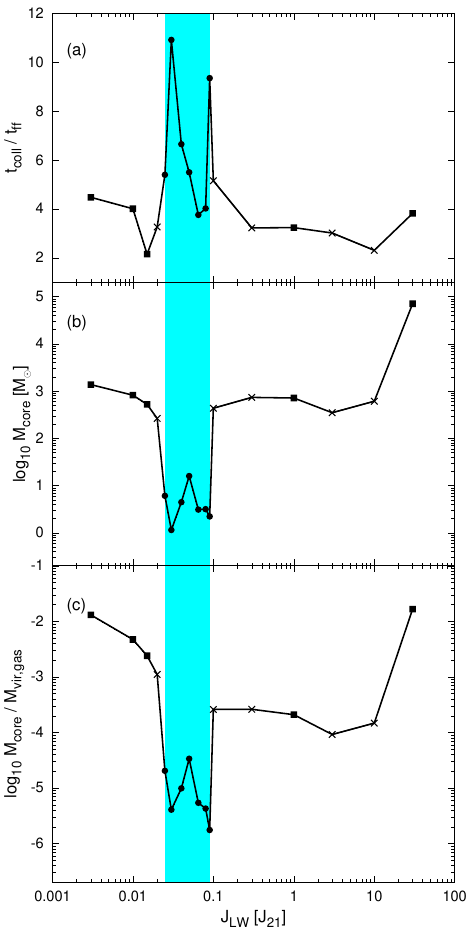}
\end{center}
\caption{
$\JLW/\Junit$ dependence of (panel a) normalized collapse timescale, (b) core mass, and (c) star formation efficiency.
The cyan region represents R3 models where the HD-cooling becomes effective ($\JLW/\Junit = 0.025 - 0.09$).
The symbols indicate whether HD-cooling is effective at two epochs ($\tth = 0$ and $10^5$\,yr), shown in columns 7-8 of Table~\ref{tab:t1}: circles, crosses, and squares mean Y-Y, Y-N, and N-N, respectively.
}
\label{fig:f6}
\end{figure}
%%%%%%%%%%%%%%%%%%%%%%%%%%%%%%%%%%%%%%%%%%%%%%%%%%%%%%%%%%%%

%%%%%%%%%%%%%%%%%%%%%%%%%%%%%%%%%%%%%%%%%%%%%%%%%%%%%%%%%%%%
\subsection{Mass of dense core} \label{sec:results:mass}
%%%%%%%%%%%%%%%%%%%%%%%%%%%%%%%%%%%%%%%%%%%%%%%%%%%%%%%%%%%%

Pop III IMF is the critical parameter to model the early universe.
This study could not directly determine the stellar mass because we artificially restrict the cloud collapse by the stiff-EOS technique.
On the other hand, we can discuss the upper limit of stellar mass by analyzing the mass of the high-density region, core mass $M_{\rm core}$ at $\nh \geq 10^8\,\cc$, inside which Pop III star(s) form.

First, all models have only one high-density core at the end of simulations.
Figure~\ref{fig:f5} plots radial profiles for all models.
During the $10^5$\,yr accretion phase, all gas clouds do not fragment regardless of increased gas mass inside the host halo from $M_{\rm vir,gas} = 1.29\times10^5\,\msun$ to $4.27\times10^6\,\msun$ with increasing LW radiation intensities from $\JLW/\Junit = 0$ to $30$.
Due to the increase in the ratio of the total gas mass to the Jeans mass, the massive gas cloud may be more likely to fragment, although no apparent Jeans-scale fragmentation occurs within the scope of the current simulations.
This is the same for HD-cooling clouds, which should more easily fragment due to the decreased Jeans mass \citep{Ripamonti2007}.

Then, we analyze the core mass for each model and summarize in Figure~\ref{fig:f6}(b).
The core mass, the upper limit of stellar mass, does not increase monotonically with external radiation intensity.
Furthermore, the core mass exhibits a characteristic behavior for classification (R1-R5):
\begin{itemize}
\item[(a)]
HD-cooling: The core masses are smaller by $1-2$ orders of magnitude, $1-16\,\msun$.
This is because the HD-cooling enabled the gas to cool further from $200$\,K and decrease the mass accretion rate as $\dot{M} \propto T^{3/2}$.
This is a new path of enabling HD-cooling by external LW radiation, resulting in low-mass Pop III star formation.
If the final mass of the star is less than $0.8\,\msun$, it has a lifetime exceeding the age of the universe (so-called ``surviving star'').
\item[(b)]
H$_2$-cooling: The core masses are about $10^3\,\msun$, independent of $\JLW$.
This is in a typical mass range for the top-heavy Pop III IMF model.
\item[(c)]
H-cooling: The core has a large mass of $7.1\times10^4\,\msun$.
It is a possible site to form an intermediate-mass black hole with $10^{4-5}\,\msun$, a possible seed for high-$z$ quasars (super-massive black holes).
\end{itemize}

%%%%%%%%%%%%%%%%%%%%%%%%%%%%%%%%%%%%%%%%%%%%%%%%%%%%%%%%%%%%
\section{Discussion} \label{sec:dis}
%%%%%%%%%%%%%%%%%%%%%%%%%%%%%%%%%%%%%%%%%%%%%%%%%%%%%%%%%%%%

%%%%%%%%%%%%%%%%%%%%%%%%%%%%%%%%%%%%%%%%%%%%%%%%%%%%%%%%%%%%
\subsection{HD-cooling clouds driven by the slow collapse} \label{sec:dis:origin}
%%%%%%%%%%%%%%%%%%%%%%%%%%%%%%%%%%%%%%%%%%%%%%%%%%%%%%%%%%%%

Our simulations show that HD-cooling is enabled in the primordial star-forming cloud exposed by the external LW radiation with intermediate intensities, $\JLW/\Junit = 0.025-0.09$ (R3 models), leading to a significant reduction in the core mass that limits the upper limit of the Pop III stellar mass.
Because HD molecules form abundantly in the ionized primordial gas, \citet{Nakauchi2014} examined the condition of efficient HD-cooling in the relic H$_{\rm II}$ region of Pop III stars irradiated by weak background LW radiation \citep[see also][]{Johnson2019}.
This work presents the condition of efficient HD-cooling in the primordial cloud without ionization by some phenomena.
In this subsection, we use the collapse timescale to explain the direct cause of HD-cooling's effectiveness and show why HD-cooling was ineffective in other models with lower and higher intensities (R2 and R4 models).

The cooling rate of HD exceeds that of H$_2$ when the abundance ratio exceeds $f_{\rm HD}/f_{\rm H2} = 10^{-3}$.
There are two main processes for the HD formation:
\begin{equation}
    \mathrm{D}+\mathrm{H}_2 \rightleftarrows \mathrm{HD}+\mathrm{H}
    \label{eq:reaction1}
\end{equation}
\begin{equation}
    \mathrm{D}^{+}+\mathrm{H}_2 \rightleftarrows \mathrm{HD}+\mathrm{H}^{+}
    \label{eq:reaction2}
\end{equation}
The reaction rate coefficients for these reactions differ significantly from those for the reverse reaction at high temperatures.
The second reaction produces more HD at low temperatures $\lesssim \! 200$\,K, however, which is the typical lower limit by H$_2$-cooling \citep{GalliPalla2002}.
There is a dilemma: for HD-cooling, which can cool the gas cloud more, to be effective, the temperature of the gas cloud must fall to just below the lower limit of the coolable temperature by H$_2$-cooling.
One solution has been proposed as a scenario in which the gas cloud collapses slowly over time, allowing the gas to lose thermal energy through prolonged cooling, thus achieving the low temperatures required for efficient HD formation \citep{Ripamonti2007}.
\cite{Hirano2014} investigated the effect of different collapse timescales on the thermal evolution of primordial gas clouds from 1-zone simulations, showing that HD-cooling is effective if the contraction timescale is about three times higher than the free-fall timescale \citep[see also][]{Gurian2023}.

We follow their argument and calculate the collapse timescale to investigate the cause of HD-cooling clouds in this study.
We define the collapse time normalized by the free-fall time, $t_{\mathrm{col}}/t_{\mathrm{ff}}$, as a collapse degree of the star-forming clouds.
The collapse time is the time it takes for the maximum gas density of the contracting gas cloud to increase by a factor of $10^{1/4}$ as
\begin{equation}
    t_{\mathrm{col}}(n) = t(n_{\rm max}=10^{1/4}n) - t(n_{\rm max}=n) \, ,
    \label{tcol}
\end{equation}
and the free-fall time is
\begin{equation}
    t_{\mathrm{ff}}(n) = \sqrt{\frac{3\pi}{32G\rho}} \simeq 5.2 \times 10^{7}\,{\rm yr} \cdot \left(\frac{n}{1\,\cc}\right)^{-1/2} \, .
    \label{tff}
\end{equation}

We calculate the normalized collapse timescale for each collapsing cloud (bottom row of Figure~\ref{fig:f4}).
Then, we determine an average value between densities $\nh = 10^{3}-10^{5}\,\cc$ at which HD formation and cooling must be effective (Table~\ref{tab:t1} and Figure~\ref{fig:f6}(a)).
Figure~\ref{fig:f6}(a) shows that HD-cooling clouds (R3 models; cyan region) experience slow contractions, $t_{\mathrm{col}}/t_{\mathrm{ff}} = 4-11$, whereas 
The collapse timescale for HD-cooling clouds is consistent with one examined in the previous work, about $10$ or more \citep[Figure~24 of][]{Hirano2014}, for primordial gas clouds with no external radiation field ($\JLW/\Junit = 0$).
Even under the LW radiation field, a sufficiently slow contraction of the gas cloud can promote HD formation.

Finally, we discuss why HD-cooling only became effective in the intermediate LW radiation intensities.
This is due to the balance between the decrease of radiative cooling efficiency by H$_2$ photodissociation and the increase of the self-gravity by the increased mass of the gas cloud with increasing radiation intensity.
First, the photodissociation rate is also low when the radiation intensity is lower (R2 models).
H$_2$ abundance recovers to the same degree as the case without radiation (R1 model) due to self-shielding effects (Figure~\ref{fig:f4}).
When H$_2$ cooling becomes effective, the gas cloud contracts while losing internal energy through H$_2$ radiative cooling.
At this time, if this energy loss rate is high, the gas cloud collapse can accelerate.
The resultant rapid contraction speed is maintained until the cloud reaches the loitering phase, leading to heating.
HD-cooling can not be effective in these cases.

On the other hand, with higher intensities (R4 models), H$_2$ are sufficiently photodissociated and H$_2$-cooling is inefficient.
However, the delay of halo formation increases the halo mass (gas mass inside the halo) before gas cloud contraction.
Due to the strong self-gravity of the cloud, the cloud can contract quickly despite the high temperature.
Then HD-cooling cannot be effective in this case either.

At intermediate intensities (R3 models), the cloud collapses over time with a reduction in cooling efficiency due to H$_2$ photodissociation, and the self-gravity of the cloud is not too strong.
Therefore, the cloud can collapse slowly, and HD-cooling is activated.

%%%%%%%%%%%%%%%%%%%%%%%%%%%%%%%%%%%%%%%%%%%%%%%%%%%%%%%%%%%%
\subsection{Temporal HD-cooling clouds} \label{sec:dis:temphdcooling}
%%%%%%%%%%%%%%%%%%%%%%%%%%%%%%%%%%%%%%%%%%%%%%%%%%%%%%%%%%%%

%%%%%%%%%%%%%%%%%%%%%%%%%%%%%%%%%%%%%%%%%%%%%%%%%%%%%%%%%%%%
\begin{figure}
\begin{center}
\includegraphics[width=1.0\linewidth]{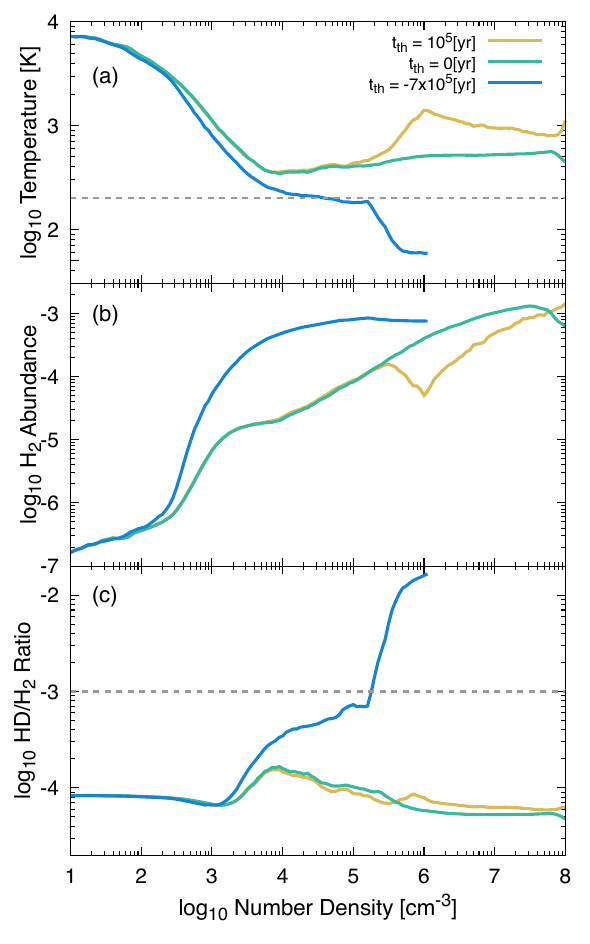}
\end{center}
\caption{
Same as the top three rows in Figure~\ref{fig:f4} for the model with $\JLW/\Junit = 0.1$.
Three lines represent results at different epochs during the simulation at $\tth = -7 \times 10^{5}$\,yr ($n_{\rm max} = 10^6\,\cc$), $\tth = 0$\,yr ($n_{\rm max} = 10^8\,\cc$), and $\tth = 10^5$\,yr.
}
\label{fig:f7}
\end{figure}
%%%%%%%%%%%%%%%%%%%%%%%%%%%%%%%%%%%%%%%%%%%%%%%%%%%%%%%%%%%%

In some models, HD-cooling that became once effective during the collapse phase ($\tth \leq 0$\,yr) became ineffective during the accretion phase ($\tth \geq 0$\,yr).
We distinguish these models as temporal HD-cooling gas clouds (Y-N combination in columns 7 and 8 of Table~\ref{tab:t1} and crosses in Figure~\ref{fig:f6}).
Figure~\ref{fig:f7} shows the thermal properties of one of the temporal HD-cooling clouds ($\JLW/\Junit = 0.1$) at three different epochs.
During the contraction phase ($\tth \leq 0$\,yr), the abundance ratio of HD and H$_2$ exceeds the critical value, and the gas temperature falls below $200$\,K in the high-density region.
However, during the accretion phase ($\tth \geq 0$\,yr), the HD-cooling becomes inefficient.
This model is classified as H$_2$-cooling cloud.

In simulation studies of star formation, stopping and analyzing simulations at the contraction stage is common practice due to computational costs.
However, simulations during the accretion phase are necessary to identify gas clouds in which HD-cooling becomes effective.

%%%%%%%%%%%%%%%%%%%%%%%%%%%%%%%%%%%%%%%%%%%%%%%%%%%%%%%%%%%%
\subsection{Restriction on the stellar mass} \label{sec:dis:mass}
%%%%%%%%%%%%%%%%%%%%%%%%%%%%%%%%%%%%%%%%%%%%%%%%%%%%%%%%%%%%

This study obtains core mass, an upper limit of the total stellar masses formed inside the core.
A part of the dense region collapses and becomes Pop III star(s) because we artificially prevent structure formation above a specific density using the stiff-EOS technique to achieve long computation times.
In addition, some physical processes occurred inside the core could cause differences between core mass and stellar mass: decrease by the circumstellar disk fragmentation \citep[e.g.,][]{Hirano2017b, Susa2019, Sugimura2020, Sugimura2023} and increase by protostellar radiative feedback from high accretion rates \citep[e.g.,][]{Hosokawa2011, Hosokawa2016, Hirano2014}.

\begin{figure}
\begin{center}
\includegraphics[width=1.0\linewidth]{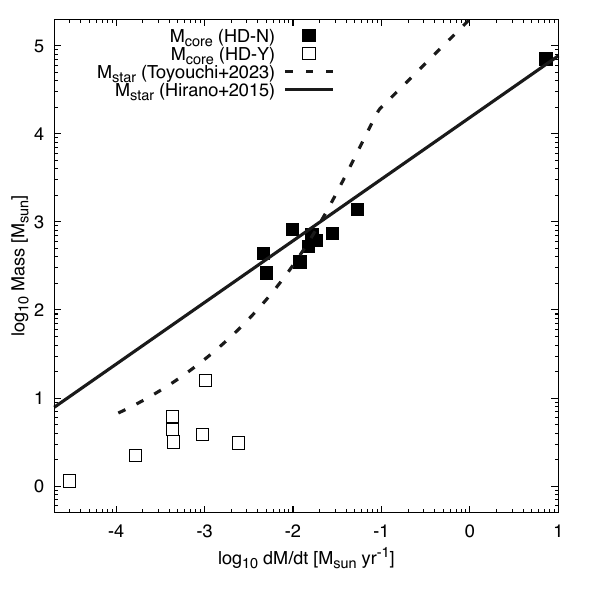}
\end{center}
\caption{
Dependence of core masses on the mass accretion rate at the Jeans radius at $\tth = 0$\,yr.
The filled and open symbols distinguish whether HD-cooling is effective at $\tth = 10^5$\,yr: without HD-cooling (filled symbols) and with HD-cooling (open).
Two lines show functions to estimate the Pop III stellar mass as a function of the mass accretion rate shown in \citet[][solid line]{Hirano2015} and \citet[][dashed]{Toyouchi2023}.
}
\label{fig:f8}
\end{figure}

The physical properties at the early phase of the star formation could determine the accretion history of the Pop III star.
Some studies constructed the correlation function to estimate the Pop III stellar mass without detail, long-term simulation to avoid huge computational cost \citep[][]{Hirano2014, Hirano2015, Toyouchi2023, Gurian2023}.
Figure~\ref{fig:f8} compares the obtained core masses and estimated stellar masses using functions in \citet{Hirano2015} and \citet{Toyouchi2023}.
The core masses of H$_2$-cooling clouds are consistent with the estimated stellar masses.
On the other hand, the core masses for HD-cooling clouds are lower than the estimated stellar mass.
This is consistent with \citet{Hirano2015}, which constructed different fitting functions for H$_2$ and HD-cooling clouds.

%%%%%%%%%%%%%%%%%%%%%%%%%%%%%%%%%%%%%%%%%%%%%%%%%%%%%%%%%%%%
%%%%%%%%%%%%%%%%%%%%%%%%%%%%%%%%%%%%%%%%%%%%%%%%%%%%%%%%%%%%

%%%%%%%%%%%%%%%%%%%%%%%%%%%%%%%%%%%%%%%%%%%%%%%%%%%%%%%%%%%%
\subsection{Restriction on the star formation efficiency} \label{sec:dis:efficiecy}
%%%%%%%%%%%%%%%%%%%%%%%%%%%%%%%%%%%%%%%%%%%%%%%%%%%%%%%%%%%%

In the context of the formation of first galaxies, one of the important targets of the JWST observations, the star formation rate is an important model parameter.
Similar to the upper limits on stellar mass shown in Section~\ref{sec:dis:mass}, we evaluate an upper limit on the star formation rate as $\fthree = M_{\rm core}/M_{\rm vir, gas} \geq M_{\mathrm{star}}/M_{\mathrm{vir, gas}}$, where $M_{\rm vir, gas}$ is the total mass of gas within the virial radius.

Figure~\ref{fig:f6}(c) shows the dependence of $\fthree$ on $\JLW$ (see also Table~\ref{tab:t1}).
The star formation efficiency decreases with radiation intensity, $\fthree \sim 10^{-2} \rightarrow 10^{-4}$.
This trend is because, with radiation intensity, the total gas mass of the halo scale increases (Table~\ref{tab:t1}) while the core mass does not vary significantly (Figure~\ref{fig:f6}(b)).
In addition, the star formation efficiency declines for HD-cooling gas clouds, $\fthree \sim 10^{-5}$, due to the decrease of core mass.
Furthermore, all models in this study form only one core, so the cloud-scale fragmentation does not contribute to increasing the total mass.

This study finds that the star formation efficiency decreases with increasing radiation intensity and even more significantly in a specific parameter range where HD-cooling is effective.
This is in direct contrast to the conventional interpretation that H$_2$ photodissociation process results in the increase of the accretion rate by increasing the gas cloud's temperature, increasing the stellar mass and star formation efficiency.
Although this study does not directly calculate the accretion process onto protostars, such a trend has been identified in the core mass, which is the upper mass limit of the accreting gas mass.

%%%%%%%%%%%%%%%%%%%%%%%%%%%%%%%%%%%%%%%%%%%%%%%%%%%%%%%%%%%%
\section{Conclusions} \label{sec:sum}
%%%%%%%%%%%%%%%%%%%%%%%%%%%%%%%%%%%%%%%%%%%%%%%%%%%%%%%%%%%%

This study examines the dependence of the Pop III star formation inside primordial gas clouds irradiated by external LW radiation with $18$ different intensities $\JLW/\Junit = 0 - 30$.
We summarize the obtained results below.

First, HD-cooling becomes effective for models with week intensities $\JLW/\Junit = 0.025-0.09$ and keeps the gas temperature below $200$\,K above a density of $10^4\,\cc$ during the first $10^5$\,yr of the protostellar accretion phase.
In HD-cooling clouds, the core mass, defined as the total gas mass of the region where $\nh \geq 10^8\,\cc$, are $M_{\rm core} = 1-16\,\msun$, which are more than two orders of magnitude smaller than in other models where HD-cooling is ineffective. 
Suppose a small core forms a low-mass Pop III star with less than $0.8\,\msun$.
In that case, the resultant star becomes an important ``fossil'' in the early Universe because its lifetime exceeds the cosmic age, and low-mass Pop III stars can survive in the Milky Way galaxy until today.
Previous studies suggested the formation of Pop III ``survivors'' from the circumstellar disk fragmentation \citep[e.g.,][]{Clark2011, Greif2011disk, Stacy2016, Wollenberg2020}.
Without considering the small-scale phenomena, we present a novel formation path of low-mass Pop III stars from cloud-scale physics.

Second, HD-cooling decreases the star formation efficiency $f_{\rm III} = M_{\rm core}/M_{\rm vir,gas}$ from $10^{-3}$ to $10^{-5}$.
The total gas mass in the halo scale increased with increasing radiation intensity while the total core mass in the high-density regions remained almost flat (Figure~\ref{fig:f6}(b)).
The core mass is independent of the radiation intensity in each of HD-/H$_2$-/H-cooling clouds.
Furthermore, regardless of the halo mass, only one high-density core is formed in all models, and we identify no cloud-scale fragmentation.

This paper examines the effect of widely different LW radiation on only one Pop III star formation site.
To generalize the effects of the weak radiation field identified in this study and determine the typical range of radiation intensity over which HD-cooling is effective, we are planning parameter survey simulations for many Pop III star-forming regions.

\begin{acknowledgments}
We thank Hisato Mori for his contributions to the early stages of this study.
Numerical computations were carried out on Cray XC50 at CfCA in National Astronomical Observatory of Japan and Yukawa-21 at YITP in Kyoto University.
Numerical analyses were in part carried out on the analysis servers at CfCA in National Astronomical Observatory of Japan.
This work was supported by JSPS KAKENHI Grant Numbers JP18H05222, JP21H01123 (S.H.) and JP21K13960 (H.U. and S.H.), Qdai-jump Research Program 02217 (S.H.), and MEXT as ``Program for Promoting Researches on the Supercomputer Fugaku'' (Structure and Evolution of the Universe Unraveled by Fusion of Simulation and AI; Grant Number JPMXP1020230406, Project ID hp230204) (S.H.).
\end{acknowledgments}

%\vspace{5mm}
%\facilities{HST(STIS), Swift(XRT and UVOT), AAVSO, CTIO:1.3m, CTIO:1.5m,CXO}

%\software{astropy \citep{2013A&A...558A..33A,2018AJ....156..123A},  
%          Cloudy \citep{2013RMxAA..49..137F}, 
%          Source Extractor \citep{1996A&AS..117..393B}
%          }

%\appendix
%\section{Appendix information}

\bibliography{ms}{}
\bibliographystyle{aasjournal}

\end{CJK}%! To show the Japanese language.
\end{document}